\documentclass{cargese}\usepackage{graphicx}

\let\footnote\savefootnote
\let\footnotetext\savefootnotetext 
\let\footnoterule\savefootnoterule 
\normallatexbib

\begin{document}
\articletitle[Learning about QCD from AdS/CFT]
{The non-AdS/non-CFT \\
correspondence, or three \\
different paths to QCD}


\author{Ofer Aharony\footnote{Incumbent of the Joseph and Celia Reskin
career development chair.}}



\affil{Department of Particle Physics \\
Weizmann Institute of Science \\
Rehovot, 76100, Israel}    

\email{Ofer.Aharony@weizmann.ac.il}

\begin{abstract}
In these lecture notes from the 2002 Cargese summer school we review
the progress that has been made towards finding a string theory for
QCD (or for pure (super)Yang-Mills theory) following the discovery of
the AdS/CFT correspondence. We start with a brief review of the
AdS/CFT correspondence and a general discussion of its application to
the construction of a string theory for QCD. We then discuss in detail
two possible paths towards a QCD string theory, one which uses a mass
deformation of the ${\cal N}=4$ super Yang-Mills theory (the
Polchinski-Strassler background) and the other using a compactification of
``little string theory'' on $S^2$ (the Maldacena-Nu\~nez solution). A
third approach (the Klebanov-Strassler solution) is described in other
lectures of this school. We briefly assess the advantages and
disadvantages of all three approaches.
\end{abstract}


\section{Introduction}

The most remarkable discovery of the last few years in theoretical
particle physics is the realization that certain string theories are
identical to certain field theories. More generally, a large class of
gravitational theories are precisely equivalent to non-gravitational
theories; we will be interested here in the special case which is the
equivalence between string theories and gauge theories.

An equivalence of this type was previously suspected on general
grounds from both directions :

{\bf Gauge theory to string theory} : Historically, string theory
originated from an attempt to describe the strong nuclear interactions,
motivated by the linear confining force between quarks and
anti-quarks (as if there is a string connecting them) and by the
appearance of Regge trajectories in the meson and hadron spectra. Now,
we know that these interactions can be described by a gauge theory
(QCD), but this gauge theory seems to behave (at least qualitatively) like a
string theory at low energies, where the running gauge coupling
constant is large. A
more quantitative reason to expect gauge theories to be related to
string theories comes from 't Hooft's analysis of the large $N$ limit
of $SU(N)$ gauge theories \cite{'tHooft:1973jz}. 
't Hooft showed that if one takes $N \to
\infty$ with fixed $\lambda_{YM} \equiv g_{YM}^2 N$, then (in a double-line
notation) the Feynman diagram expansion arranges itself according to
the minimal genus $g$ of the surface which the diagrams can be drawn on,
and every amplitude can be written in the form $\sum_{g=0}^{\infty}
N^{2-2g} f_g(\lambda_{YM})$ with some functions $f_g$ arising from genus
$g$ diagrams. This is the same expansion as in a string theory whose
string coupling constant is $g_s \simeq 1/N$. The leading contribution
in the large $N$ limit comes from ``planar diagrams'' with $g=0$.
't Hooft conjectured that for
large $\lambda_{YM}$ the Feynman diagrams become dense and form smooth two
dimensional surfaces which may be identified with string worldsheets.

{\bf Gravity to field theory} : String theory is a theory of quantum
gravity. It was argued by 't Hooft and Susskind (see Bousso's lectures
at this school) that the number of degrees of freedom in a theory of
quantum gravity on some manifold scales as the area of the boundary of
the manifold rather than as the volume of the manifold. This suggests
that theories of quantum gravity may be described as non-gravitational theories
living on the boundary of space-time; in some cases this description
could involve local field theories.

The first explicit realization of these general expectations was the
AdS/CFT correspondence, and we will start by briefly reviewing this
correspondence. The correspondence allows us to use string theory and
gravity to learn about the strong coupling limits of specific field
theories, such as the ${\cal N}=4$ supersymmetric Yang-Mills (SYM)
theory. The standard model includes in it QCD, which is an $SU(3)$
gauge theory which becomes strongly coupled at low energies, and it
would be very interesting to find a good description of this theory in
the regime where the original variables of QCD (the quarks and gluons)
become strongly coupled. As discussed above, one expects such a
description to involve string theory (at least for large $N$, but the
stringy description could perhaps be useful even for $N=3$). In these
lectures we will discuss how we may be able to get such a description
from generalizations of the AdS/CFT correspondence, in several
different ways. All these ways will lead to a description of the large
$N$ limit of a pure Yang-Mills theory (sometimes with ${\cal N}=1$
supersymmetry), but this is presumably a necessary first step towards
constructing a string theory for QCD, and the theories we will discuss
have the same qualitative features as QCD (exhibiting phenomena like
confinement and chiral symmetry breaking). We will encounter some
obstacles on all the different paths towards a string theory of QCD,
but the obstacles seem to be technical rather than problems of
principle, so hopefully it will be possible to overcome them in the
future.

\section[Review of AdS/CFT]{A brief review of the AdS/CFT\\ correspondence}

The first concrete example (above two dimensions) to be found of the
general relation between large $N$ gauge theories and string theories
is the AdS/CFT correspondence \cite{juan,gkp,wittenads}.  In general
this relates a theory of quantum gravity on $d+1$ dimensional anti-de
Sitter (AdS) space (times some compact manifold) to a $d$-dimensional
conformal field theory (CFT). In the case of type IIB string theory
compactified on 5-dimensional AdS space $AdS_5$ (times a compact
5-dimensional manifold) the dual theory is usually a gauge theory. 

The simplest example of the AdS/CFT correspondence is the duality
between type IIB string theory compactified on $AdS_5\times S^5$ and
the ${\cal N}=4$ supersymmetric four dimensional Yang-Mills theory
with gauge group $SU(N)$. The parameters of string theory in this
background\footnote{It is often said that string theory has no
continuous parameters since all its apparent parameters are actually
fluctuating fields. For instance, the string coupling constant is
related to the vacuum expectation value of the dilaton field. However,
while this is true in flat space, it is not true in backgrounds like
AdS space, where the fluctuations of the asymptotic value of fields
like the dilaton are non-normalizable, so the asymptotic values are
frozen and serve as parameters for the string theory.} are the string
coupling constant $g_s$ and the radius of curvature $R$ of the $AdS_5$
and of the $S^5$, measured in units of the string tension $1 / 2\pi
\alpha^{\prime}$; from these two parameters one can determine (using
the equations of motion) the (quantized) flux of the 5-form field of
type IIB string theory on the $S^5$. The parameters of the Yang-Mills
theory are the Yang-Mills coupling constant $g_{YM}$ (which does not
depend on the scale in this theory) and the number of colors $N$. The
two sets of parameters are related by
\begin{equation}
\label{relone}
4\pi g_s = g_{YM}^2
\end{equation}
and
\begin{equation}
\label{reltwo}
R^4 / (\alpha^{\prime})^2 = g_{YM}^2 N.
\end{equation}
These two relations imply that the flux of the 5-form field on the $S^5$
is equal to $N$ (in the units in which it is quantized to be an
integer).

The AdS/CFT correspondence is believed to be an exact equivalence, in
the sense that all observables are supposed to be equal on both sides
(when the parameters are identified according to equations
(\ref{relone}), (\ref{reltwo})). Some of the evidence for this
equivalence is reviewed in \cite{magoo}. It is often said that the
AdS/CFT correspondence is a strong/weak coupling duality (like
S-duality), but this is not precise since each side of the
correspondence is really labeled by two parameters. As described
above, at the leading order in $1/N$ we keep only planar diagrams in
the field theory, and these can be described by a perturbation theory in
the parameter $\lambda_{YM} = g_{YM}^2 N$; on the other hand at leading
order in $g_s$ in the string theory, we have in the string worldsheet
sigma model an expansion in the space-time curvature $\alpha^{\prime}
/ R^2$. Equation (\ref{reltwo}) tells us that these two expansions are
indeed related by a strong/weak coupling duality. On the other hand,
the large $N$ genus expansion in the field theory is governed by
$1/N$, and the string theory genus expansion is governed by $g_s$, and
equation (\ref{relone}) tells us that for fixed $\lambda_{YM}$ these two
expansions are actually the same. Thus, if we could somehow perform
exact computations in $\lambda_{YM}$ in the field theory, or in
$\alpha^{\prime}$ in the string theory, then the AdS/CFT correspondence
would turn into a perturbative duality where we could compare results
from both sides order by order in $g_s$ (or in $1/N$). An example of
this was recently found in the plane-wave limit of $AdS_5\times S^5$
\cite{bmn}, where the string theory worldsheet sigma model is exactly
solvable \cite{metsaev}.

The matching of all observables related to local operators can be
formulated via the equivalence of the partition function of the
conformal field theory with the partition function of string
theory. We use the metric $ds^2 = (dz^2 + d{\vec x}^2) / z^2$ for AdS
space\footnote{These coordinates cover the Euclidean AdS space if we
choose a Euclidean signature for the ${\vec x}$ coordinates, or a
Poincar\'e patch of the Lorentzian AdS space if we choose a Lorentzian
signature for the ${\vec x}$ coordinates.}, in which the boundary of
the space, spanned by the ${\vec x}$ coordinates, is at $z=0$. If we
denote by ${\cal O}_i(x)$ the gauge-invariant local operators in the
CFT, with scaling dimension $\Delta_i$, and by $\chi_i(x,z)$ the
corresponding string theory fields in AdS space, whose mass is related
to the operator dimensions by
\begin{equation}
\label{massdim}
R^2 m_{\chi_i}^2 = \Delta_i (\Delta_i - 4), 
\end{equation}
then we can write the
equivalence as
\begin{equation}
\label{equivalence}
Z[\lambda_i] \equiv \langle e^{i \int d^4x \lambda_i(x) {\cal O}_i(x)}
\rangle_{CFT} = Z_{IIB}[\lim_{z \to 0}(\chi_i(x,z) z^{\Delta_i-4}) = 
\lambda_i(x)].
\end{equation}
Here we introduced arbitrary sources $\lambda_i(x)$ for every
operator, and on the right-hand side we have the string theory
partition function with particular boundary conditions on the boundary
of AdS space.  The correlation functions of local operators in the
theory are given by derivatives of this expression with respect to
$\lambda_i(x)$ (at $\lambda_i=0$).

In fact, the relation (\ref{equivalence}) only holds for a particular
class of operators called ``single-trace operators'', which in the
gauge theory may be written as a trace of a product of adjoint
representation fields; the generalization of this formula to include
also ``multiple-trace operators'' which cannot be written in this way
is discussed in \cite{abs,wittenmulti,bss} and is less well-understood
on the string theory side.

A particular interesting class of operators in the ${\cal N}=4$ SYM
theory includes the chiral primary operators, which are in short
representations of the ${\cal N}=4$ superconformal algebra such that their
dimensions cannot receive any quantum corrections. Some of these are
mapped to type IIB supergravity fields, and for large field theory
coupling $\lambda_{YM}$ we can compute their correlation functions on
the right-hand side of (\ref{equivalence}) in the supergravity
approximation (since $\alpha^{\prime} / R^2$ is small). In this
approximation, the boundary conditions corresponding to putting in
delta-function sources for the fields (as we usually do in computing
correlation functions) are imposed by bulk-to-boundary propagators in
the supergravity computation. More generally, we have in the string
theory (at least at weak string coupling) a vertex operator
$V_i(x,\sigma)$ for each single-trace operator ${\cal O}_i(x)$ (where
$\sigma$ is a coordinate on the string worldsheet), and we can
identify ${\cal O}_i(x)$ with $\int d^2 \sigma V_i(x,\sigma)$ in the
sense that inserting the former into the field theory path integral is
the same as inserting the latter into the string theory path
integral. In this way we can map all single-trace local observables
between the two sides of the correspondence.

This mapping of operators is precisely known only for the chiral
primary operators (and for some related states with large R-charges,
as described in Minwalla's lectures). Denoting the field content of
the ${\cal N}=4$ SYM theory by $A_{\mu}$ for the gauge field,
$\lambda^a$ ($a=1,2,3,4$) for the fermions and $\phi^i$
($i=1,2,3,4,5,6$) for the scalar fields (all these fields are in the
adjoint representation of $SU(N)$), the chiral primary operators of
the ${\cal N}=4$ superconformal algebra are given by ${\rm
tr}(\phi^{i_1} \phi^{i_2} \cdots \phi^{i_k})$ ($k=2,3,\cdots,N$) where
the $SO(6)_R$ indices $i_j$ are contracted in a symmetric traceless
manner. All other single-trace chiral operators in the theory are
descendants of these (they can be obtained from the operators above by
acting on them with the supercharges of the theory), and their
spectrum agrees precisely (for $k\ll N$) with that of the
single-particle states of type IIB
supergravity compactified on $AdS_5\times S^5$. This precise matching
means that all other operators in the SYM theory must be matched to
other fields of type IIB string theory which are not type IIB
supergravity fields. All these fields have a mass at least of the
order of the string scale. Using (\ref{massdim}) and (\ref{reltwo})
this means that the dimension of the corresponding operators grows at
least as fast as $(g_{YM}^2 N)^{1/4}$ in the limit of large $N$ and
large $g_{YM}^2 N$.

Many tests and generalizations of the correspondence have been made in
the last five years, which I will not go into here -- some of them are
described in \cite{magoo} (where more details of the correspondence
may also be found).

\section[From AdS/CFT to QCD]{From AdS/CFT to QCD-like\\ theories}
\label{toqcd}

As described above, the AdS/CFT correspondence allows us (among other
things) to learn about the strong coupling behavior of the ${\cal
N}=4$ SYM theory (for instance, to compute the anomalous dimensions of
certain operators in the strong coupling limit) from type IIB string
theory (or supergravity). This field theory is quite different from
QCD -- in particular it is scale-invariant (both classically and
quantum mechanically), while QCD is weakly coupled at high energies
but strongly coupled at low energies. For describing nature it would
be more interesting to understand various strong coupling properties
of QCD, such as confinement, chiral symmetry breaking, the formation
of a mass gap, the spectrum of bound states like mesons and baryons,
and so on. All these properties are strong coupling effects that
cannot be computed in QCD perturbation theory. However, as described
in the introduction above, there is no reason why it shouldn't be
possible to learn about these properties by using a dual string theory
(with weak string coupling in the large $N$ limit), analogous to the
one appearing in the AdS/CFT correspondence reviewed in the previous
section.

In the previous section we saw that the string theory dual of the
${\cal N}=4$ SYM theory had a limit where it was well approximated by
supergravity. On the other hand, a string theory background describing
QCD (or pure Yang-Mills theory) must be strongly curved. One argument
for this is that asymptotically free theories like QCD are weakly
coupled at high energy scales (compared to some characteristic scale
$\Lambda_{QCD}$), and we saw above that weak coupling tends to be
related to large curvatures; in particular it is hard to imagine how
we could reproduce perturbative QCD amplitudes at high energies from a
supergravity theory. A more general argument is that the spectrum of
QCD-like theories, including gauge-invariant particles like mesons and
glueballs, comes (approximately) in Regge trajectories (recall that
this was one of the original reasons to suspect that they are related
to string theories). The Regge trajectory including particles of spin
$J$ starts at a mass of order $M_J \simeq \sqrt{J} \Lambda_{QCD}$. On
the other hand, a weakly curved superstring theory background has a
supergravity approximation at low energies, and supergravity theories
include only particles of spin less than or equal to 2, so in such
theories there is a large mass gap between the spectrum of particles
with $J \leq 2$ and with $J > 2$ (the low-energy theory is generally a
ten dimensional supergravity theory, which gives rise via a
Kaluza-Klein reduction to many particles of spin 2 or lower in four
dimensions).

Thus, it seems that to approach QCD we will eventually need to
understand string theory in highly curved backgrounds -- this has been
one of the main obstacles to formulating a string theory of QCD. As
described above, the coupling between strings is not an obstacle,
since in the large $N$ limit we expect to find a dual string
theory with a string coupling constant of order $g_s \simeq 1/N$ (but
this could also become a problem when we eventually take $N=3$).

Unfortunately, progress in studying strongly curved backgrounds in
string theory has been rather slow (except for some special cases,
like WZW models, which are well understood). Thus, so far almost all
computations in string theory duals to QCD-like theories have been done in
supergravity approximations. As described above, such approximations
will never be quantitatively similar to QCD. However, they can be
qualitatively similar to QCD, and can exhibit similar properties like
confinement and chiral symmetry breaking. As we will see, they could
even be continuously related to theories like pure Yang-Mills theory
by changing parameters -- for some range of parameters of the theory
we could have a good supergravity approximation (and have a large
ratio between $M_{J=3}$ and $M_{J=2}$) while in a different range of
parameters the same theory could go over to pure Yang-Mills theory
(with a small ratio between $M_{J=3}$ and $M_{J=2}$). Thus, it is
interesting to study such ``toy models of QCD'' even if we do not
currently know how to control their behavior in the regime where they
approach QCD, and this will be the main subject of these lecture
notes.

Above we described the AdS/CFT correspondence for the case of the
${\cal N}=4$ SYM theory. In order to approach QCD we want to find
string theory duals for
non-conformal theories which have less (or no)
supersymmetry. There are three general ways that have been used so far to
obtain such theories from the AdS/CFT correspondence : \begin{itemize}
\item{(i)} We can start from a conformal theory (with a known string
  theory dual) and deform it by a relevant or marginal operator which
  breaks conformal invariance and supersymmetry.
\item{(ii)} We can start from a higher dimensional theory (again with
  a known string theory dual) and compactify it to four dimensions in
  a way which (partially) breaks supersymmetry.
\item{(iii)} We can try to directly find the dual of non-conformal
  four dimensional field theories.
\end{itemize}
The last method leads to ``duality cascades'', and it is described in
  Klebanov's talks in this school. In these talks I will focus on the
  first two methods in sections (\ref{deform}) and
  (\ref{compactification}), respectively, and briefly discuss the third
  method in section (\ref{klestrass}).

\section[Deformations of N=4 SYM]{Deformations of the 
${\cal N}=4$ SYM\\ theory}
\label{deform}

\subsection{Deformations in field theory} 

Since the ${\cal N}=4$ SYM theory provides the simplest example of the
AdS/CFT correspondence, the simplest way to use the first method is to
study deformations of this theory. The ${\cal N}=4$ theory includes
four massless adjoint fermions $\lambda^a$ and six massless adjoint
scalars $\phi^i$, and
we would like to get rid of them in order to remain with the pure
Yang-Mills theory. As mentioned above, this theory is already quite
similar to QCD. Adding a small number of quark flavors is not
expected to significantly change the large $N$ limit of the theory,
but it adds some complications, so we will be content here
with finding a dual for pure Yang-Mills theory. 

So, we would like to deform the ${\cal N}=4$ theory by mass terms for
the ``extra'' fields, doing something like
\begin{equation}
\label{naivedef}
{\cal L}_{SYM} \to {\cal L}_{SYM} + {\rm
tr}(M^2 \phi^i \phi^i + ({\tilde m}_{ab} \lambda^a \lambda^b + c.c.)). 
\end{equation}
The
problem with this deformation is that the first term we wrote down,
${\rm tr}(\phi^i \phi^i)$, is not a chiral operator (it is the
``trace'' part of ${\rm tr}(\phi^i \phi^j)$, and it sits in the
so-called Konishi multiplet). This not only means that we cannot study
it using supergravity (since it is not in the supergravity spectrum,
which includes only chiral operators),
but also that for large $g_{YM}^2 N$ this operator acquires a large
anomalous dimension (as described above), and the deformation by this
operator becomes irrelevant (so it does not make sense as a quantum
field theory). We can write down chiral scalar mass operators of the
form $(M^2)_{ij} {\rm tr}(\phi^i \phi^j)$ with traceless $(M^2)_{ij}$,
but then the sum of the
eigenvalues of $(M^2)_{ij}$ must vanish, so some of the eigenvalues
must be negative, leading to instabilities and to a breaking of the gauge
symmetry (which we do not desire).

Thus, the only mass deformation which makes sense as a leading
deformation is of the form $({\tilde m}_{ab} {\rm tr}(\lambda^a
\lambda^b) + c.c.)$. The mass matrix can always be diagonalized so we
can generally write this as 
\begin{equation}
\label{fermmass}
m_1 {\rm tr}(\lambda^1 \lambda^1) + m_2
{\rm tr}(\lambda^2 \lambda^2) + m_3 {\rm tr}(\lambda^3 \lambda^3) +
m_4 {\rm tr}(\lambda^4 \lambda^4) + c.c. 
\end{equation}
This involves chiral operators ${\rm tr}(\lambda^a \lambda^b)$
(arising from the action of two supercharges on the chiral primary
operator ${\rm tr}(\phi^i \phi^j - {1\over 6} \delta^{ij} \phi^k
\phi_k)$), of (protected) scaling dimension $\Delta=3$, so we can
study this deformation (at least in principle) in the supergravity
approximation. We expect that when we perform such a deformation
scalar masses will be induced quantum mechanically (by loop
corrections, by operator mixings or by RG flow) at order $m^2$, and we
could also turn on tree-level scalar masses of order $m^2$
proportional to the chiral scalar mass operator. But, we should still
require that the masses squared of all the scalars are positive so
that we have a theory which does not spontaneously break the gauge
symmetry.

One way to ensure this stability property is to retain some of the
supersymmetry of the original theory. This requires that one of the
fermions must remain massless, $m_4=0$, so that it can be in the same
supersymmetry multiplet as the gauge field. If $m_4=0$ we can always choose
the second order scalar masses such that the theory preserves
supersymmetry, and then we can describe the deformation by a
superpotential (in ${\cal N}=1$ superspace notation). In ${\cal N}=1$
language the ${\cal N}=4$ theory includes one vector multiplet and
three chiral multiplets $\Phi_i$ in the adjoint representation, and we
can describe the theory after the mass deformation by the
superpotential
\begin{equation}
\label{massdeform}
W = {1 \over {g_{YM}^2}} {\rm tr}(2 \sqrt{2} \Phi_1 [\Phi_2, \Phi_3] +
m_1 \Phi_1^2 + m_2 \Phi_2^2 + m_3 \Phi_3^2).
\end{equation}
At leading order in the deformation this is exactly the fermion mass
term we wrote above (with $m_4=0$, where we normalize the kinetic
terms of all the fields to be proportional to $1/g_{YM}^2$), and at
second order the theory with the superpotential (\ref{massdeform})
includes particular scalar mass terms which ensure that an ${\cal
N}=1$ subgroup of the original ${\cal N}=4$ supersymmetry remains
unbroken. This ensures that even after we take quantum corrections
into account, the scalar masses squared will always be
non-negative. If we do not preserve supersymmetry this may or may not
happen\footnote{After these talks were given the case of four equal
fermion masses was analyzed in \cite{bce}, and some evidence was given
for the claim that this case does not give rise to any tachyonic
scalar modes. However, the solution presented there appears to be
singular.}.

One simple possibility is to give a supersymmetry-preserving mass only
to a single chiral multiplet by taking only $m_1$ to be non-zero. In
this case the mass deformation is believed to lead at low energies to
a strongly coupled superconformal field theory
\cite{Leigh:1995ep,Karch:1999pv} with a superpotential of the form $W
\sim \lambda {\rm tr}([\Phi_2,\Phi_3]^2)$ (obtained by integrating out
$\Phi_1$ in (\ref{massdeform})). The supergravity dual of this theory
is known (though the solution describing the full mass deformation has
only been computed numerically and not explicitly
\cite{Khavaev:1998fb,Pilch:2000fu}), and $\lambda$ is an exactly
marginal operator in the low-energy conformal theory which maps to the
dilaton in its string theory dual\footnote{Note that this is an
example of a theory where having small $g_s N$ in the string theory
does not lead to a weakly coupled theory on the field theory side of
the duality.}. If we turn on only two fermion masses (one way to do
this is to choose $|m_1|=|m_2|$, in which case the deformation
actually preserves ${\cal N}=2$ supersymmetry) we find at low energies
a theory with a Coulomb branch along which a vacuum expectation value
(VEV) for $\Phi_3$ breaks the gauge symmetry to $U(1)^{N-1}$. The
supergravity dual of this deformation is known at some special points
in its moduli space (called ``enhan\c con points'')
\cite{Johnson:1999qt,Pilch:2000ue, Buchel:2000cn,Evans:2000ct}. Both
for one non-zero mass and for two non-zero masses the low-energy
behavior is very different from that of QCD so we will not discuss it
here.

The interesting case for us will be the case where all three of the
supersymmetry-preserving masses are non-zero. Naively this gives a
mass to all the fields except for the ${\cal N}=1$ vector multiplet,
so for energies $E \ll m_i$ we expect to remain with the pure ${\cal
N}=1$ SYM theory. This theory is quite similar to QCD -- it also confines,
exhibits a mass gap of the order of a strong coupling scale
$\Lambda_{SYM}$ and chiral symmetry breaking, and so
on. Unfortunately, this naive expectation is not always realized. We
can easily compute the characteristic scale $\Lambda_{SYM}$ of this
``low-energy'' SYM theory, defined by $\Lambda_{SYM}^b = \mu^b \, {\rm
exp}(-8\pi^2 / g_{YM}^2(\mu))$ where $\mu$ is the renormalization
scale and $b$ is the one-loop beta function, by matching the running
coupling constant at the mass scales of the chiral multiplets. We find
$\Lambda_{SYM}^3 = m_1 m_2 m_3\, {\rm exp}(-8\pi^2 / g_{YM}^2 N)$
where $g_{YM}^2$ is the gauge coupling constant of the high-energy
${\cal N}=4$ SYM theory. This means that for large $g_{YM}^2 N$, where
we might expect to have a dual supergravity description (at least we
know that we have one for energies above the mass scales),
$\Lambda_{SYM}$ is of the same order as the $m_i$, so there is no
separation of scales between the fields $\Phi_i$ that we are trying to
get rid of and the states of the ``low-energy'' ${\cal N}=1$ SYM
theory, since the lowest mass states of this theory are expected to
have masses of order $\Lambda_{SYM}$. In order to get such a
separation we must take $g_{YM}^2 N \ll 1$. This leads to
$\Lambda_{SYM} \ll m_i$, but then we do not have a good supergravity
approximation and we must deal with a strongly curved string theory
background. This is in agreement with our general discussion in the
previous section.

When all  three masses are non-zero  it turns out that  the theory has
many discrete  supersymmetric vacua. Classically the  vacua are given,
as in  any other  supersymmetric gauge theory,  by solving  the F-term
equations $dW  / d\Phi_i = 0$  and dividing the space  of solutions by
the complexified gauge  group, which in our case  is $SL(N,C)$ (this
ensures that the D-term equations are also satisfied). Taking all the
masses to be equal for simplicity (the generalization to arbitrary
masses is straightforward), the F-term equations are 
\begin{equation}\label{fterm}
[\Phi_i, \Phi_j] = -{m\over \sqrt{2}} \epsilon_{ijk} \Phi_k.
\end{equation}
Since this is the defining equation for a representation of the
$SU(2)$ Lie algebra
(up to a rescaling of the $\Phi_i$'s), this is solved by any three 
matrices which are an $N$-dimensional representation of $SU(2)$. Any
such representation is a sum of irreducible representations of
dimensions $n_j$, and we can write the classical solutions for the
matrices $\Phi_i$ in a block-diagonal form involving blocks of size
$n_j\times n_j$ (with $\sum_j n_j = N$). Only three of the six real
matrices $\phi^i$ are non-zero in these vacua, and the choice of
which three is determined by the phase of the mass $m$.

In most of these vacua the gauge group is not completely broken, so
classically there is no mass gap. The only vacuum where the gauge
group is completely broken classically is the one corresponding to the
$N$-dimensional representation of $SU(2)$, which is called the ``Higgs
vacuum''. The opposite case is when we have $N$ copies of the
$1$-dimensional representation of $SU(2)$, and then $\langle \Phi_i
\rangle = 0$. Classically this leaves the full $SU(N)$ group
unbroken. Quantum mechanically we expect in this vacuum to get at low
energies a behavior like that of the pure ${\cal N}=1$ SYM theory (as
described above, this expectation is valid for small $g_{YM}^2 N$),
which confines (so this is called the ``confining vacuum'') and has
$N$ different vacua (differentiated by the ``gluino condensate''
$\langle {\rm tr}(\lambda \lambda) \rangle$). An intermediate case is
the vacuum corresponding to $p$ copies of the $q$-dimensional
representation of $SU(2)$ (with $pq=N$). Classically in such a vacuum
the gauge group is broken to $SU(p)$, and the low-energy gauge theory
is the ${\cal N}=1$ SYM theory with gauge group $SU(p)$, so quantum
mechanically we expect this vacuum to have a mass gap and split into
$p$ discrete vacua \cite{Donagi:1995cf}. All other vacua, involving irreducible
representations of different dimensions, have classically unbroken
$U(1)$ gauge groups, so we expect that they will not have a mass gap
even in the quantum theory (since the $U(1)$ vector multiplets, which
are free at low energies, are expected to remain massless).

In field theory, much is known about the behavior of this theory in
its different vacua. For example, one can exactly compute the vacuum
expectation values of various chiral operators like ${\rm tr}(\Phi^k)$
and ${\rm tr}(\lambda \lambda)$
\cite{Dorey:1999sj,Dorey:2000fc,Aharony:2000nt,Dorey:2002ad}. In the
non-supersymmetric case we expect to get a similar qualitative
behavior of the theory (if the deformation is stable), except that the
degeneracy between the different vacua is expected to be lifted (it is
not clear whether the ``confining vacua'', which are related to the pure
Yang-Mills theory, are the ones with the lowest vacuum energy).

\subsection{Deformations in string theory}

What is the string theory dual of the mass-deformed field theory in
all the different vacua described above ? In order to compute the
correlation functions of the ${\cal N}=4$ SYM theory one uses the
relation (\ref{equivalence}) only for infinitesimal
$\lambda_i(x)$. However, this relation is valid also for finite
$\lambda_i(x)$. Thus, deformations of the ${\cal N}=4$ SYM theory by
$\int d^4 x \lambda_i(x) {\cal O}_i(x)$ correspond to solutions of
string theory with an asymptotically AdS background in which $\langle
\chi_i(x,z) \rangle \simeq \lambda_i(x) z^{\Delta_i-4}$, or
equivalently by adding to the worldsheet action a term of the form
$\int d^4 x \lambda_i(x) \int d^2 \sigma V_i(x,\sigma)$ (instead of
just inserting this combination into correlation functions). We are
interested in Lorentz-invariant deformations for which $\lambda_i(x)$
is independent of $x$.

In the supergravity approximation, in order to describe a particular
deformation we need to find a solution to type IIB supergravity with
the appropriate boundary conditions. For supersymmetric solutions
(which we are mainly interested in here) this can actually be done by
solving only first order equations (second order equations are
generally needed for non-supersymmetric solutions).  When we decompose
the ten dimensional type IIB supergravity multiplet on $AdS_5\times
S^5$ into multiplets of
five dimensional ${\cal N}=8$ supergravity, the field $\chi$ appearing
in the mass deformation belongs to the lowest multiplet, which is the
five dimensional graviton multiplet (containing also the massless
graviton). It is believed that in order to find solutions involving
the fields in this multiplet one can first solve the five dimensional
supergravity equations of motion and then lift the solution to ten
dimensions (this is called ``consistent truncation''), and solving
these equations is not very difficult. However, it turns out
\cite{gppz} that doing this leads to a family of singular solutions
(which become singular at the interior of AdS space), parameterized by
different (continuous) values for $\langle {\rm tr}(\lambda \lambda)
\rangle$ (where $\lambda$ is the adjoint fermion which classically
remains massless), while from the field theory discussion above we
expect to find a discrete family of solutions, with particular known
values for $\langle {\rm tr}(\lambda \lambda) \rangle$ in each
discrete vacuum. How can we resolve this mismatch ? Since we find
singular solutions it is clear that the supergravity approximation
breaks down, so we have to go beyond this approximation.

Some hints for how we should go beyond supergravity are provided by
the following facts :
\begin{itemize}
\item{(i)} The supergravity field $\chi$ corresponding to the mass
  operator ${\rm tr}(\lambda^a \lambda^a)$, which gets a vacuum
  expectation value when we perform the mass deformation, is a
  particular linear combination of the components of the NS-NS and R-R
  2-form fields with indices in the $S^5$. For small values of $z$
  (near the boundary) this expectation value grows as $\chi \propto z$
  as we move into the interior of AdS space, suggesting that in order
  to desingularize the solution we may need to insert a source for
  these 2-form fields (though the total charge of the 3-form field
  strengths vanishes, so there is no necessity for doing this). Such a
  source would be a D5-brane or an NS5-brane wrapped on an $S^2$
  inside the $S^5$, filling the four dimensional space $R^4$ which the
  gauge theory lives on, and localized in the radial coordinate $z$.
\item{(ii)} In the ``Higgs vacuum'' the gauge group is completely
  broken classically, and (at weak coupling) we can choose this
  breaking to occur at a scale much larger than the strong coupling
  scale $\Lambda_{SYM}$ of the theory, so we expect quantum
  corrections in this vacuum to be small. This vacuum corresponds to
  the VEVs of the $\Phi_i$ being given by the $N$-dimensional
  irreducible representation of $SU(2)$. In this representation we
  have $\sum_i |\Phi_i|^2 \simeq |m|^2 N^2$ (as a matrix equation,
  with the identity matrix implied on the right-hand side), so it
  looks like the eigenvalues of the VEVs of the $\Phi_i$ sit on a
  sphere in the moduli space with radius $|m| N$ (recall that only
  three of the six real fields $\phi^i$ are non-vanishing, so this is
  an $S^2$ and not an $S^5$). For large $N$ the eigenvalues almost
  completely cover this sphere, though it should be noted that since
  the $\Phi_i$ do not commute this geometrical description is slightly
  naive (the vacuum expectation values really sit on a ``fuzzy
  sphere'' and not on a sphere). Now, if we think of the $\Phi_i$'s as
  corresponding to the positions of D3-branes (which is how they
  started life in the ``derivation'' of the AdS/CFT correspondence
  before we took the near-horizon limit), then such vacuum expectation
  values correspond to a configuration where the $N$ D3-branes are
  blown up into a spherical D5-brane (this is often called ``Myers'
  dielectric effect'' \cite{Kabat:1997im,Myers:1999ps}). This suggests
  again that (at least) the ``Higgs vacuum'' could be related to a
  spherical D5-brane wrapped on an $S^2$ inside the $S^5$.
\end{itemize}

The arguments above motivate looking for solutions which are not just
described by supergravity but which involve also 5-branes wrapped on
2-spheres inside the $S^5$. Polchinski and Strassler \cite{polstr}
found that non-singular solutions of this type indeed exist with the
appropriate boundary conditions. The solutions are not known
explicitly, but their form is known asymptotically near the boundary
of AdS and near the 5-branes, and the 5-branes are stable in these
backgrounds (and even exhibit a classical mass gap if there is only a
single 5-brane) even though topologically they could collapse to a
point since they carry no total 5-brane charge. It is believed that
these asymptotic solutions can be pieced together into non-singular
backgrounds of supergravity including 5-branes (which may be viewed as
specific sources for the supergravity fields). I will not describe
here the details of these solutions. A special case of the
Polchinski-Strassler solutions involves a single D5-brane, so it is
natural to identify it with the ``Higgs vacuum'' of the gauge theory
following the discussion above. This D5-brane carries $N$ units of
gauge field flux on the $S^2$, $\int_{S^2} F \simeq N$, which implies
(using the $F \wedge C_4$ coupling in the Wess-Zumino term of the
D5-brane action) that it carries $N$ units of D3-brane charge. Thus,
one might say (motivated by the discussion above) that the original
$N$ D3-branes have blown up into a D5-brane in this configuration,
though this is somewhat misleading since the D3-branes were not there
in $AdS_5\times S^5$ before we did the deformation\footnote{One could try to
study these deformations in the full D3-brane background before taking
the near-horizon limit, but it is not known how to do this.}. The
D5-brane sits at some particular value of $z_0$ where it is stable
($z_0$ is proportional to $1 / \alpha^{\prime} m N$), and since it carries
$N$ units of charge of the R-R 5-form field $F_5$, this means that the total
3-brane charge $\int_{S^5} F_5$, which equals $N$ near the boundary of
AdS, decreases as we increase $z$. For $z \gg z_0$ there is no
longer any 5-form flux, and the solution in fact goes over to flat
space there. So, this solution interpolates between $AdS_5\times S^5$
for small $z$ and $R^{10}$ for large $z$ (with a 5-brane in the
middle).

Similar solutions were found to exist also with $p$ D5-branes, each of
which carries $q$ units of D3-brane charge (with $pq=N$) and sitting
at $z_0 \simeq 1 / \alpha^{\prime} m q$. It is natural to identify these
with the vacua of the gauge theory in which the gauge symmetry is
classically broken to $SU(p)$ -- in fact we can explicitly see this
$SU(p)$ as the classical gauge symmetry on the (overlapping)
D5-branes. Both in the field theory and in the string theory this
$SU(p)$ gauge group is expected to confine, and each such vacuum is
expected to split into $p$ distinct vacua, but this is not visible in
the classical string theory.

The solutions described above are weakly coupled everywhere if we have
$g_s p \ll 1$, and they are weakly curved far away from the D5-branes
for $g_s N \gg 1$ (the regions near the D5-branes can be described
well by open+closed string theory as usual). If we continue the logic
of the previous paragraph, we would find that the $\langle \Phi
\rangle =0$ ``confining'' vacuum should be described by $N$ coincident
D5-branes. However, such a collection of D5-branes would have a large
back-reaction on the background (since $g_s N$ is assumed to be large)
and lead to large curvatures, so we cannot use supergravity to analyze
it or even to check if such a solution exists or not. 

Luckily, by using S-duality we can find a different description of the
same vacuum (more precisely, of one of the $N$ vacua coming from the
classical $\langle \Phi \rangle = 0$ vacuum). From a direct field
theory analysis one can argue that the $SL(2,Z)$ S-duality symmetry of
the ${\cal N}=4$ SYM theory permutes the different vacua, and in
particular the S-duality transformation $g_{YM} \to 4\pi /g_{YM}$ (for
zero theta angle) exchanges the ``Higgs vacuum'' (where the $\Phi_i$
which are electrically charged degrees of freedom condense, so we
expect magnetically charged degrees of freedom to confine as in the
Meissner effect in superconductors) with one of the ``confining''
$\langle \Phi \rangle =0$ vacua (where we expect the electrically
charged particles to be confined, and magnetically charged particles
to condense) \cite{Donagi:1995cf}. So, we can try to describe the
$\langle \Phi \rangle =0$ vacuum by performing an S-duality
transformation on the solution that we found for the ``Higgs
vacuum''. Recall that the AdS/CFT correspondence maps the
electric-magnetic S-duality of the ${\cal N}=4$ SYM theory to the
S-duality of the type IIB string theory. The S-duality transformation
of the ``Higgs vacuum'' gives a background with a single NS5-brane and
no D5-branes, and it turns out that such a solution indeed exists and
gives a good description for the ``confining vacuum'', in which (for
large $N$ and $g_s N$) supergravity and string perturbation theory are
good approximations except in the vicinity of the
NS5-brane. Similarly, other 5-brane configurations can be found to
describe all the other vacua of the mass-deformed theory. Whenever we
find more than one possible description for a vacuum (as we found
above for the ``confining vacuum''), they do not have an overlapping
range of validity, so at most one solution can be trusted for
describing a particular vacuum at a particular value of the coupling
constant.

The description of the various vacua in terms of branes wrapping
$S^2$'s was recently confirmed directly in the field theory by Dorey
and Sinkovics in \cite{Dorey:2002ad}. They computed the vacuum
expectation values of the chiral operators ${\rm tr}(\Phi^k)$ in the
various vacua at large $g_{YM}^2 N$ (this is possible since exact
expressions are known for the VEVs of chiral operators in the
mass-deformed SYM theory), extracted
from them the distributions of the eigenvalues of the $\Phi_i$
matrices, and showed that indeed they lie on spheres in the range of
parameters where we have solutions in which the 5-form charge is carried
by 5-branes wrapping spheres.

While the field theory analysis we made relied strongly on
supersymmetry, the construction of the solutions described above does
not depend on supersymmetry, and one can find qualitatively similar
solutions also in some non-supersymmetric cases, when all four fermion
masses are non-zero. At least when the supersymmetry-breaking mass is
small, one can argue that these solutions are still stable, since the
original supersymmetric solutions had a mass gap (see below), and
adding a small additional mass deformation cannot destabilize the
background when the new mass is much smaller than the mass gap. Of
course, supersymmetry breaking generically removes the degeneracy
between the different vacua, so we expect that most of the solutions
of \cite{polstr} should be meta-stable (classically stable but quantum
mechanically unstable) after a small supersymmetry breaking
deformation, and one of them should be stable.

\subsection[Comparison with QCD]{QCD-like properties of the\\ 
Polchinski-Strassler theory}

As we discussed above, the regime where the solution of \cite{polstr}
is well-described by supergravity is quite different from the regime
where it is equivalent (at low energies) to a pure ${\cal N}=1$ SYM
theory. However, the two regimes are continuously related (by changing
$g_s$) so one expects to find the same qualitative properties in this
background as one has in SYM, and this is indeed the case.

First, it is believed that the SYM theory dynamically generates a mass
gap of the order of the QCD scale $\Lambda_{SYM}$. One can show that
the background of \cite{polstr} also has a mass gap. For the bulk
fields this requires solving their equations of motion (in the
presence of the 5-branes), looking for non-singular solutions whose
dependence on the $R^4$ directions is of the form $e^{ik\cdot x}$, and
showing that the spectrum of the $3+1$ dimensional $m^2=k^2$ is
strictly positive -- one can argue that this is indeed the case in the
solution of \cite{polstr} even without knowing the explicit
solution. Similarly, one has to solve the equations of motion for the
fields living on the branes and show that they also have a mass
gap. In the background with $p$ overlapping 5-branes we classically
have a $U(p)$ gauge theory in $5+1$ dimensions living on the 5-branes,
compactified on $R^4 \times S^2$. It turns out that the $U(1)$ factor
is actually unphysical (it can be gauged away; when there is more than
one group of overlapping 5-branes only the sum of their $U(1)$ fields
may be gauged away), so one is left with an $SU(p)$ gauge theory, and
one can show that the classical massless spectrum on the 5-branes is
that of a $3+1$ dimensional ${\cal N}=1$ super Yang-Mills theory with
gauge group $SU(p)$. In particular, for the ``confining vacuum'' and
the ``Higgs vacuum'' where $p=1$ there are already classically no
massless fields on the 5-branes, so the full background has a mass
gap. For the vacua with $p > 1$ we expect that quantum effects on the
5-branes will generate a mass gap for the $SU(p)$ gauge theory just
like they do on the gauge theory side of the correspondence, so we
will have a mass gap also in this case but it is not directly visible
in the supergravity (plus 5-branes) approximation. In vacua with more
than one group of overlapping 5-branes some $U(1)$ gauge multiplets
remain massless, and these configurations correspond to the vacua
without a mass gap in the field theory.

Another qualitative property of the quantum SYM theory is confinement
of fields charged (electrically) under the gauge group, which is
expected to be related to condensation and screening of magnetically
charged particles. As discussed above, we expect to see this property
in the ``confining vacua'', while we expect to see screening of
electrically charged particles in the ``Higgs vacuum''. This property
is usually tested by putting in an external quark-anti-quark pair and
computing the force between them (or, equivalently, the Wilson loop
operator). It is easy to see that in the ``Higgs vacuum'' one indeed gets
screening. External electrically charged particles are described by
fundamental strings ending on the boundary of (asymptotically) AdS
space \cite{Maldacena:1998im,Rey:1998ik}, and these strings can end on
the D5-brane, so the force between charges at different positions in
$R^4$ decays at least like the force between charged particles in a
$3+1$ dimensional $U(1)$ gauge theory. On the other hand, in the
``confining vacuum'' whose description involves an NS5-brane, the string
cannot end anywhere, but must stretch between the external particles,
so the potential energy turns out to be proportional to the distance
between the external particles as expected for confinement. For
magnetically charged sources, which are described by D-strings ending
on the boundary, the situation is the opposite, as expected from
general considerations. The behavior in other vacua is also as
expected \cite{Kinar:2000wh}.

Other QCD-like properties of the solution of \cite{polstr} were also analyzed,
such as the existence of a baryon vertex, instantons, condensates
analogous to the gluino condensate of the pure SYM theory, etc.

Let us end this section by summarizing the advantages and
disadvantages of this solution as a string theory for QCD. In the
limit $g_s N \to 0$ the solution provides (at energies much smaller
than $m$) a string theory for pure SYM theory (or, when we turn on
supersymmetry-breaking masses, for pure YM theory), as desired. As
expected from general considerations, this limit involves large
curvatures. In the case of \cite{polstr} it also involves a rather
complicated sigma model, including RR fields, whose form is not known
exactly even in the supergravity approximation. Moreover, in the case
of the ``confining vacuum'' (which is most closely related to the pure SYM
theory) the solution involves NS5-branes, which are not under control
in perturbative string theory. So, while these solutions provide
implicit string theories for pure SYM theory, it seems very difficult
to analyze them in the limit which is interesting for QCD. On the plus
side, the construction of \cite{polstr} easily generalizes to the
non-supersymmetric case, so if one understands the supersymmetric case
(and uses it to construct a string theory for pure SYM theory) it should be
possible to easily find also a string theory for the
(phenomenologically more interesting) pure Yang-Mills theory.

\section{Compactification scenarios}
\label{compactification}

Two main compactification scenarios have been extensively discussed in
the literature (though many others are also possible). Both of them
start from maximally supersymmetric configurations, such that only the
compactification breaks supersymmetry to four dimensional ${\cal N}=1$
or ${\cal N}=0$ supersymmetry. The first scenario \cite{Witten:1998zw}
involves the six dimensional conformal field theory, arising as the
low-energy limit on M5-branes, compactified on $S^1\times S^1$, such
that the fermions have anti-periodic boundary conditions around one of
the circles. In an appropriate decoupling limit this gives the pure
non-supersymmetric $3+1$ dimensional Yang-Mills theory. The second scenario
\cite{Maldacena:2000yy} involves the ``little string theory'', coming
from the decoupling limit of type IIB NS5-branes, compactified on an
$S^2$ inside a Calabi-Yau manifold. In an appropriate decoupling limit
this gives the $3+1$ dimensional ${\cal N}=1$ SYM theory. We will
focus on the second scenario here, because it has the advantage (which
we did not have in the previous section) of having a limit where it is
weakly coupled and weakly curved everywhere (with no branes), and it
does not involve any Ramond-Ramond fields, so in this limit this
scenario is amenable to standard string theory computations. However,
a disadvantage of this scenario is that the high-energy theory is much
more complicated than in the previous section (where it was just the
${\cal N}=4$ SYM theory).

\subsection{``Little string theories''}

Let us begin by reviewing the decoupling limit of NS5-branes in
general, before they are compactified. Since in type IIB string theory
NS5-branes are S-dual to D5-branes, their decoupling limit should be
the same. In general, to obtain a decoupled theory on D-branes we
need to take the Planck length $l_p$ to zero so as to remove the
gravitational interactions with the bulk modes, but we want to keep
the brane Yang-Mills coupling constant $g_{YM}$ fixed so as to remain
with a non-trivial theory on the brane. Recall that $l_p^8 \propto g_s^2
(\alpha^{\prime})^4$, and that for Dp-branes $g_{YM}^2 \propto g_s
(\alpha^{\prime})^{(p-3)/2}$. Thus, for $p = 3$ the decoupling limit
takes $\alpha^{\prime} \to 0$ and keeps $g_s$ constant, for $p < 3$
we need to take both $g_s$ and $\alpha^{\prime}$ to zero while keeping
$g_{YM}$ constant, while for $p=4,5$ we need to take $g_s$ to infinity
and $\alpha^{\prime}$ to zero, again keeping $g_{YM}$ constant (there
is no decoupling limit for $p > 5$). Since we are taking $g_s$ to be
very large we cannot trust the original string theory description, but
in the type IIB case we can use an S-dual description; 
note that S-duality does
not change $l_p$ and $g_{YM}$ (which are physically measurable by
low-energy interactions). After the S-duality we find for $p=5$ that the same
decoupling limit can be described by starting with $N$ NS5-branes 
in type IIB string theory and taking $g_s \to 0$ while
keeping $\alpha^{\prime}$ constant (in this S-dual description; note
that on NS5-branes $g_{YM}^2 \simeq \alpha^{\prime}$).

This limit \cite{Berkooz:1997cq} is called a ``little string theory''
(LST) since it involves a string tension parameter $\alpha^{\prime}$
which is kept constant in the decoupling limit but it does not involve
a massless graviton; see \cite{Aharony:1999ks} for a review of these
theories. The resulting decoupled theories are not local field
theories. For instance, they have (at high energies) a Hagedorn-like
density of states, $S \propto E$, as in free string theories, they
have a continuous spectrum of single-particle states (above a mass gap
$M_0 \simeq 1 / \sqrt{N \alpha^{\prime}}$), and when compactified on
circles they have a T-duality symmetry. All these properties are most
easily seen from the holographic dual of the ``little string
theories'' \cite{Aharony:1998ub}. This is just the near-horizon limit
of the string theory background of $N$ NS5-branes
\cite{Callan:1991at}. The metric of this near-horizon limit in the
string frame is simply
\begin{equation}
\label{nsmetric}
ds^2 = dx_{R^{5,1}}^2 + N \alpha^{\prime} d\rho^2 + N \alpha^{\prime} 
ds^2_{S^3},
\end{equation}
and there is also a linear dilaton in the $\rho$ direction, $g_s = g_0 
e^{-\rho}$, and $N$ units of NS-NS 3-form flux on the $S^3$ (such that
  the sigma model on the $S^3$ is simply the $SU(2)$ Wess-Zumino-Witten
 model at level $N$).

From the S-duality to D5-branes it is clear that at low energies the
``little string theory'' reduces to a $5+1$ dimensional SYM theory
(with ${\cal N}=(1,1)$ supersymmetry) with gauge group $SU(N)$ and
with coupling constant $g_{YM}^2 \propto \alpha^{\prime}$. This theory
is not renormalizable so it requires some high-energy completion,
which is provided by the LST. The holographic description is
problematic because the string coupling diverges as we take $\rho \to
-\infty$. It is not surprising that the holographic description is not
under full control, since the low-energy theory (at energies below $1/
\sqrt{N \alpha^{\prime}}$) is weakly coupled, and it is hard to imagine
how weakly coupled SYM computations could emerge from a simple string
theory background. This problem will be less severe after we
compactify (there are also ways to get around it in six dimensions).

\subsection[Compactified LSTs]{Compactified ``little string\\ theories''}

Next, we want to reduce the ``little string theory'' to a $3+1$
dimensional field theory. Compactifying on a torus $T^2$ would give at
low energies an ${\cal N}=4$ theory in $3+1$ dimensions, so in order
to reduce the amount of supersymmetry we need to compactify on an
$S^2$; when we compactify NS5-branes on a 2-cycle with this topology
which is embedded inside a Calabi-Yau manifold the configuration
preserves precisely $3+1$ dimensional ${\cal N}=1$ supersymmetry. At
low energies we find the $5+1$ dimensional SYM theory compactified on
an $S^2$, with a particular embedding of the spin connection of the
$S^2$ inside the $SO(4)$ R-symmetry of the $5+1$ dimensional SYM
theory (which follows from the way that the $S^2$ is embedded in the
Calabi-Yau manifold). This turns out to give a mass of the order of
the inverse radius of the $S^2$ to the four scalar fields in the $5+1$
dimensional vector multiplets, and to most of the fermionic fields as
well. Classically, the only remaining massless degrees of freedom are
those of a $3+1$ dimensional ${\cal N}=1$ SYM theory (namely, $SU(N)$
gauge fields and a fermion field in the adjoint representation). There
are two scales where additional degrees of freedom appear. One is the
scale $1 / R_{S^2}$ which comes from the Kaluza-Klein reduction on the
$S^2$, and the other is the scale $1 / \sqrt{N \alpha^{\prime}}$ where
additional degrees of freedom of the LST become important (beyond its
massless modes discussed above).

By the usual matching of coupling constants, the four
dimensional coupling constant at the scale $1/R_{S^2}$ is
$g_{YM}^2(1/R_{S^2}) \simeq {\alpha^{\prime}} / R_{S^2}^2$.
Below this scale we can use the running coupling of the ${\cal N}=1$
SYM theory, and we find that the strong coupling scale of the theory
is given by
\begin{equation}
\Lambda_{SYM} \simeq {\exp(-{{8 \pi^2 R_{S^2}^2} / {3 N
      \alpha^{\prime}}})\over R_{S^2}}.
\end{equation}
In order to get a decoupling of the SYM theory from all the other
degrees of freedom we need $\Lambda_{SYM} \ll 1 / R_{S^2}$ and
$\Lambda_{SYM} \ll 1 / \sqrt{N \alpha^{\prime}}$, and we see that this
requires $N \alpha^{\prime} \ll R_{S^2}^2$, so the radius of the
sphere should obey $R_{S^2} \gg
\sqrt{N \alpha^{\prime}}$. In this limit we obtain at low energies (of
order $\Lambda_{SYM}$) the
pure SYM theory. However, from the general considerations we described in
section (\ref{toqcd}) we expect that this limit will not correspond to a weakly
coupled and weakly curved string theory dual. We will see that this is
indeed the case.

\subsection{The string theory dual}

The discussion above suggests that the compactified theory has one
dimensionless parameter, $R_{S^2}^2 / \alpha^{\prime}$, in addition to
$N$ and to the string scale $\alpha^{\prime}$ (which were parameters
already in the six dimensional theory). The holographic dual of this
compactification
was found (as a solution to supergravity, which should correspond to a
solution to type IIB string theory as well) by Maldacena and Nu\~nez
in \cite{Maldacena:2000yy}, by modifying and reinterpreting a
solution from \cite{Chamseddine:1997nm}. 

To write the solution, let us parameterize the $S^3$ of
(\ref{nsmetric}) by $\psi, {\tilde \theta}$ and $\phi$, and define a
basis of 1-forms by $w^1 + i w^2 = e^{-i\psi} (d{\tilde \theta} + i \sin
(\tilde \theta) d\phi)$, $w^3 = d\psi + \cos (\tilde \theta) d\phi$ such
that 
the volume form on $S^3$ is $w^1 \wedge w^2 \wedge w^3$. Let us
parameterize the $S^2$ by $ds_{S^2}^2 = d\theta^2 + \sin^2(\theta)
d\varphi^2$; note that these coordinates parameterize the $S^2$ which the
dual ``little string theory'' is compactified on, but in our ten dimensional
solution it is not necessarily the minimal 2-cycle at every value of the
radial coordinate\footnote{This issue was discussed most clearly in
\cite{Bertolini:2002yr} which was published after these lectures were
given.}. And, let us
define the three components of an $SU(2)$ gauge field on
$S^2$ by 
\begin{equation}
A^1 = {1\over 2}a(\rho) d\theta,\quad A^2 = {1\over 2}a(\rho)
\sin(\theta) d\varphi,\quad A^3 = {1\over 2} \cos(\theta) d\varphi, 
\end{equation}
where
$a(\rho) \equiv 2 \rho / \sinh(2\rho)$, and denote the corresponding
$SU(2)$ field strength by $F^a$. Then, the solution involves the
string frame metric
\begin{equation}
\label{mnmetric}
ds_{str}^2 = dx_{R^{3,1}}^2 + N \alpha^{\prime} 
[d\rho^2 + e^{2g(\rho)} (d\theta^2 +
  \sin^2(\theta) d\varphi^2) + {1\over 4} \sum_a (w^a - A^a)^2],
\end{equation}
where
\begin{equation}
e^{2g(\rho)} = \rho \coth(2\rho) - {\rho^2 \over \sinh^2(2\rho)} -
{1\over 4}.
\end{equation}
The dilaton is given by
\begin{equation}
g_s^2 = e^{2h_0} {2e^{g(\rho)}\over \sinh(2\rho)},
\end{equation}
and the NS-NS 3-form is given by
\begin{equation}
\label{mnthreeform}
H = N [-{1\over 4} (w^1 - A^1) \wedge (w^2 - A^2) \wedge (w^3 - A^3) +
  {1\over 4} \sum_a F^a \wedge (w^a - A^a)].
\end{equation}

Asymptotically, at large values of $\rho$, the solution is similar to
the six dimensional case (\ref{nsmetric}) but with an $R^4\times S^2$
replacing the $R^6$, and with the volume of the $S^2$ growing linearly
with the radial coordinate $\rho$. As $\rho \to 0$, a 2-cycle
shrinks smoothly to zero size, while the size of the $S^3$ remains
constant. So, the region near $\rho=0$ looks like $R^7 \times S^3$,
and the full solution (\ref{mnmetric}) interpolates between $R^7\times
S^3$ near $\rho=0$ and $R^4\times S^2\times S^3\times R_{linear\
dilaton}$ for large $\rho$. The dilaton, which decreases linearly as
$\rho \to \infty$, goes to a constant value $g_s = e^{h_0}$ at the
minimal radial position $\rho=0$, which is the maximal value of the
string coupling in this solution. This maximal string coupling is the
additional dimensionless parameter of this solution (compared to the
six dimensional solution (\ref{nsmetric})). By examining the
asymptotic value of the volume of the $S^2$, one can show that $h_0
\propto R_{S^2}^2 / N \alpha^{\prime}$. Thus, as expected, when this
ratio is large (as required for the SYM limit) we have strong coupling
at small values of $\rho$, and we do not have a good string theory
description of the limit where we get the SYM theory\footnote{For very
strong coupling we can do an S-duality transformation and study the
S-dual background, but then the string coupling diverges for large
$\rho$ and there is an intermediate regime where the string coupling
is of order one so neither description is valid. For large enough
$h_0$ the supergravity approximation also breaks down and large
curvatures arise for small $\rho$.}. However, for small $h_0$ the
solution has weak coupling, weak curvatures and no Ramond-Ramond
fields, so it is an ``ideal'' holographic background in which
computations can easily be performed in string theory (or in a
supergravity approximation), and it is continuously related to the
pure SYM theory.

\subsection[Comparison to QCD]{QCD-like properties of the\\ 
Maldacena-Nu\~nez theory}

Even in the regime of small $h_0$, where we can trust supergravity
(and weakly coupled string theory), we can see that the theory has
many properties similar to the ones we expect in the pure SYM theory :
\begin{itemize}
\item{} Mass gap -- we can analyze the low-energy spectrum using
  supergravity and see that the theory has a mass gap. Of course, in
  the regime where we can trust supergravity the QCD scale
  $\Lambda_{SYM}$ is of the same order as the Kaluza-Klein scale, and
  the mass gap is also of this order, so it does not make sense to
  compare the quantitative results to the SYM theory.
\item{} Confinement -- we can compute the quark-anti-quark force at
  large distances and see that the potential it is linear. In this
  case, since we are talking about a gauge theory originating from
  NS5-branes, the external charged particles are the ends of
  D-strings, so the tension of the confining string (the ``QCD
  string'') is that of a D-string sitting at $\rho=0$ and stretching
  in the $R^{3,1}$ directions. For large $h_0$ we can use the S-dual
  description for small $\rho$, and in that regime the confining string tension
  is that of a fundamental string sitting at $\rho=0$ in the S-dual
  background.
\item{} Chiral symmetry breaking -- one can show that the asymptotic
  metric at large $\rho$ has a $U(1)_R$ symmetry which can be
  identified with the $U(1)_R$ symmetry rotating the gluino in the
  pure SYM theory \cite{Maldacena:2000yy,Gomis:2001xw}. In field theory this
  symmetry is broken by the chiral anomaly to $Z_{2N}$, and it is then
  spontaneously broken further to $Z_2$ by the gluino condensate
  $\langle {\rm tr}(\lambda \lambda) \rangle$. In supergravity, this
  symmetry breaking turns out to be visible already at the classical
  level \cite{Klebanov:2002gr,Bertolini:2002xu}. 
  In the asymptotic region of large
  $\rho$ the $U(1)_R$ symmetry is spontaneously broken to $Z_{2N}$ by
  the 3-form field (\ref{mnthreeform}), and in the region of small
  $\rho$ this symmetry is broken further to $Z_2$.
\item{} Many other QCD-like properties are also visible in this
  background, like instantons, baryons, domain walls, the gluino
  condensate, and so on
  \cite{Maldacena:2000yy,Loewy:2001pq,Apreda:2001qb}, but we
  will not discuss them in detail here.
\end{itemize}

\section{The Klebanov-Strassler background}
\label{klestrass}

A third type of background which leads in an appropriate limit to the
pure SYM theory was presented in \cite{Klebanov:2000hb}. I will not
discuss it in detail here, since it is discussed in detail in
Klebanov's talks at this school. I will focus here only on the
similarities and differences between this approach to QCD and the
other approaches which were described in detail above.

Like the Maldacena-Nu\~nez background described in the previous
section, the solution of \cite{Klebanov:2000hb} also gives a smooth
supergravity background, it has a parameter which can be continuously
varied to give (in some limit) the pure ${\cal N}=1$ SYM theory
(though of course the supergravity approximation breaks down in this
limit), and it has the same nice QCD-like properties discussed in the
previous subsection (including chiral symmetry breaking). An advantage
of the solution of \cite{Klebanov:2000hb} is that one does not
encounter strong string coupling in the decoupling limit, but only
large curvatures (unlike the previous case, and like the
Polchinski-Strassler case if we ignore the presence of the NS5-brane
there). An important
difference between the solutions it that the solution of
\cite{Klebanov:2000hb} involves Ramond-Ramond fields, so it is more
difficult to analyze on the worldsheet; however, note that as
discussed above, if we want to use the solution of
\cite{Maldacena:2000yy} in the regime where it approaches QCD then
we need to use S-duality and then 
this solution also has RR fields. Another difference is that the
high-energy properties of the Klebanov-Strassler solution are not
completely clear -- it appears to correspond to a new type of
high-energy behavior which had not previously been encountered, called
a ``duality cascade'' in \cite{Klebanov:2000hb}. Unlike the previous
case that we discussed, here the theory remains $3+1$ dimensional at high
energies, but the number of degrees of freedom appears to grow
logarithmically with
the energy scale.

\section{Summary}

At this point (summer 2002) we have several holographic duals to
theories which at low energies include $3+1$ dimensional
$SU(N)$ Yang-Mills theory (or the ${\cal N}=1$ SYM theory), and
involve also various additional massive fields. All of these dual
theories have a parameter which interpolates between a regime where we
can control the string theory side of the duality (we have weak
coupling and weak curvatures) and another regime where the (S)YM
fields decouple from the massive degrees of freedom. However, the
qualitative properties of the theory do not depend on this parameter
-- in the supersymmetric case one can argue that no phase transitions
occur as this parameter is changed, and it is likely that this is true
also in some non-supersymmetric cases. So, one can verify various
qualitative properties of Yang-Mills theory using these holographic
duals, such as confinement, the formation of a mass gap, chiral
symmetry breaking, etc. On the other hand, quantitative properties of
the theory, like the ratio between the confining string tension and
the square of the mass gap, depend non-trivially on the
interpolating parameter, so they cannot be computed without obtaining
some control over string theory in the strongly curved (and
sometimes also strongly coupled) regime.

These constructions give us a proof in principle that (super)
Yang-Mills theory has a dual string theory description, involving the
usual type IIB string theory (rather than exotic new string theories),
and formally
we can define this dual string theory by a limiting procedure. However,
currently we do not know how to control this limit, or how to directly
define the limiting string theory without all the extra degrees of
freedom. Of course, we expect that the three different paths described
above should all lead to the same string theory in the appropriate
decoupling limit, but it is not clear how this actually happens
(perhaps there are several dual descriptions of the limiting string
theory). 

In these talks I focussed on the supersymmetric case. Going to
non-supersymmetric Yang-Mills theory seems to be straightforward in
the first scenario I described (the Polchinski-Strassler background),
but it is more difficult in the other two scenarios. In the scenario
of \cite{Maldacena:2000yy} one can perhaps do this by
supersymmetry-breaking deformations along the lines of
\cite{Aharony:2002vp}; in the scenario of \cite{Klebanov:2000hb} this
may also be possible but the rules for performing deformations there
are not completely clear since we do not fully understand the
high-energy theory. I discussed here only the case of $SU(N)$ gauge
theories, but it is not difficult to generalize all the solutions
found above also to the case of $SO(N)$ and $USp(N)$ gauge groups
\cite{polstr,Aharony:2000cw,Gomis:2001xw,Klebanov:2000hb}.
I also focussed in these talks exclusively on the
pure (super)Yang-Mills case, although for QCD we obviously want to add
quark flavors as well. A recent general discussion on some possible
ways to do this may be found in \cite{Karch:2002sh}.

\begin{acknowledgments}

It is a pleasure to thank all the people I collaborated with on
related topics, and all the people with which I discussed large $N$
field theories and string theory, which are too numerous to be listed
here. I would like to thank the organizers of the Carg\`ese 2002
Advanced Summer Institute for an enjoyable and stimulating school, and
the students at the school for asking many good questions during these
lectures.  This work was supported in part
by the Israel-U.S. Binational Science Foundation, by the ISF Centers
of Excellence Program, by the European RTN network HPRN-CT-2000-00122,
and by the Minerva foundation.

\end{acknowledgments}

\end{document}